\documentclass[11pt,a4paper]{article}

\usepackage{jcappub}
\usepackage[latin1]{inputenc}
\usepackage{bbm}
\usepackage{graphicx}
\usepackage{epsfig}
\usepackage{color}
\usepackage{slashed}

\usepackage{graphicx}
\usepackage{caption}
\usepackage{subcaption}
\usepackage{verbatim}

\allowdisplaybreaks

\newcommand{\be}{\begin{equation}} 
\newcommand{\ee}{\end{equation}}
\newcommand{\bea}{\begin{equation}\begin{aligned}} 
\newcommand{\eea}{\end{aligned}\end{equation}}

\newcommand{\bmp}{\noindent\begin{minipage}{16cm}}
\newcommand{\emp}{\end{minipage}\vskip 7mm} 
\def\lsim{\mathrel{\raise.3ex\hbox{$<$\kern-.75em\lower1ex\hbox{$\sim$}}}}
\def\gsim{\mathrel{\raise.3ex\hbox{$>$\kern-.75em\lower1ex\hbox{$\sim$}}}}

\newcommand{\intron}[1]{}

\newcommand{\GeV}{\,{\mathrm{GeV}}}
\newcommand{\MeV}{\,{\mathrm{MeV}}}

\renewcommand{\i}{\mathrm{i}}

\def\muphis{\mu_{\mathit{\phi s}}} 
\def\lphis{\lambda_{\hspace{-0.2mm}\mathit{\phi s}}}

\subheader{\small{preprint: HIP-2015-15/TH}}

\title{Self-interacting dark matter and cosmology of a light scalar mediator}
\author[a,c]{Kimmo Kainulainen,}
\author[b,c]{Kimmo Tuominen} 
\author[a,c]{and Ville Vaskonen}

\affiliation[a]{Department of Physics, University of Jyv\"askyl\"a, \\
                      P.O.Box 35 (YFL), FI-40014 University of Jyv\"askyl\"a, Finland}
\affiliation[b]{Department of Physics, University of Helsinki, \\
                      P.O.Box 64, FI-00014 University of Helsinki, Finland}                 
\affiliation[c]{Helsinki Institute of Physics, \\
                      P.O.Box 64, FI-00014 University of Helsinki, Finland}
\emailAdd{kimmo.kainulainen@jyu.fi}
\emailAdd{kimmo.i.tuominen@helsinki.fi}
\emailAdd{ville.vaskonen@jyu.fi}

\abstract{We consider a fermionic dark matter candidate interacting via a scalar mediator coupled with the Standard Model through a Higgs portal. We consider general setting including both scalar and pseudoscalar interactions between the scalar and fermion, and illustrate the relevant features for dark matter abundance, direct search limits and collider constraints. The case where dark matter has a self-interaction strength $\left\langle \sigma_V \right\rangle/m_\psi \sim 0.1-1 \,\mathrm{cm}^2/\mathrm{g}$ is strongly constrained, in particular by the Big Bang Nucleosynthesis. We show that these constraints can be alleviated by introducing a new light sterile neutrino $N$. The allowed region for the extended model consists of a triangle at $10 (\sin \theta)^{-2/5} \,{\rm MeV} \lsim m_N \lsim 1\GeV$.}


\begin{document}
\maketitle

%
\section{Introduction}
%

Recently there has been a lot of interest in models with extended matter sectors very weakly coupled with the Standard Model (SM) fields, see e.g. \cite{McDonald:1993ex, Burgess:2000yq,LopezHonorez:2006gr,Alvares:2012qv,LopezHonorez:2012kv, Fairbairn:2013uta,Alves:2013tqa,Alanne:2014bra,Hambye:2008bq, Davoudiasl:2013jma, DiChiara:2015bua}. These are mainly motivated by the need to explain the dominating dark matter (DM) component of the energy density of the Universe and simultaneously hiding the beyond SM fields from the direct searches at collider experiments. Alternatively, the extra matter fields coupled with the SM scalar sector also affect the properties of the electroweak phase transition and enhance the possibility of successful electroweak baryogenesis.

The simplest possibility of a hidden sector consists of a single singlet scalar coupled with the SM via the Higgs portal, see e.g. \cite{McDonald:1993ex,McDonald:2001vt,Cline:2012hg,Cline:2013gha} for  DM studies and \cite{Profumo:2007wc, Espinosa:2011ax} for aspects of electroweak phase transition. While the singlet scalar can in principle act as DM and lead to  strong first order phase transition, in light of present constraints for the DM abundance and from the direct search experiments, these features cannot be realized simultaneously \cite{Cline:2013gha}. To alleviate this issue one is led to consider more complex hidden sector structures. The possibilities of singlet fermions and vectors have been considered in the literature \cite{LopezHonorez:2012kv, Fairbairn:2013uta,Alves:2013tqa,Alanne:2014bra,Hambye:2008bq, Davoudiasl:2013jma, DiChiara:2015bua}. For example in the case of singlet fermion DM interacting with the SM via Higgs portal one can easily provide sufficient DM abundance and simultaneously use the scalar mediator to provide for strongly first order electroweak phase transition \cite{Alanne:2014bra}.

Further motivations for extended hidden sectors arise from the possibility of DM self-interactions as has been studied in various contexts \cite{Spergel:1999mh,Wandelt:2000ad,Faraggi:2000pv,Mohapatra:2001sx,Kusenko:2001vu,Loeb:2010gj,Kouvaris:2011gb,Rocha:2012jg,Peter:2012jh,Vogelsberger:2012sa,Zavala:2012us,Kahlhoefer:2013dca,Tulin:2013teo,Kaplinghat:2013xca,Kaplinghat:2013yxa,Cline:2013pca,Cline:2013zca,Petraki:2014uza,Buckley:2014hja,Boddy:2014yra,Schutz:2014nka}. Such interactions would allow to reconcile the tensions between the observations and simulations of small-scale structure of collisionless cold DM (CCDM): First, as the observations of dwarf galaxies confirm, the DM density profile is not cusped towards the center of the galaxy as seen in simulations of CCDM  \cite{Navarro:1996gj} but exhibits a flat core \cite{Moore:1994yx,Flores:1994gz}. Second, the so-called missing satellite problem \cite{Klypin:1999uc,Moore:1999nt,Kauffmann:1993gv} and the too big to fail problem \cite{BoylanKolchin:2011de} can also be explained by sizable interactions between the DM particles, although alternative astrophysical explanations also exist \cite{Liu:2010tn,Tollerud:2011wt,Strigari:2011ps,Oh:2010mc,Brook:2011nz,Pontzen:2011ty,Governato:2012fa}. 

In this paper we consider a simple paradigm for self-interacting DM: a singlet fermion $\psi$ interacting via scalar mediator $s$. We consider in particular the parameter region where the scalar mediator $s$ is light. In \cite{Loeb:2010gj} it has been pointed out that self-interactions of a DM particle $\psi$ mediated with a light force carrier $s$ and satisfying $(m_\psi/10$ GeV$)(m_s$/100 MeV$)^2\sim 1$, can provide the flat density profile around the core of dwarf galaxies for a wide range of interaction strengths between $\psi$ and $s$ while evading constraints on self-interactions from galactic and cluster scales. However, there exist additional stringent constraints. Most notably, if the scalar $s$ couples to the SM via the Higgs portal \cite{Burgess:2000yq,Patt:2006fw,Andreas:2008xy,Andreas:2010dz,Djouadi:2011aa,Pospelov:2011yp,Greljo:2013wja,Bhattacherjee:2013jca}, it must decay before the onset of the big bang nucleosynthesis (BBN) around $t\sim 1$ sec. in order that the success of the BBN explaining the abundances of light elements is not endangered. This sets a lower bound on the strength of coupling between the hidden and visible sectors which is too strong \cite{Kaplinghat:2013yxa} in light of present constraints for the invisible Higgs decay at LHC and direct search constraints from the LUX experiment.  Generic ways to evade this problem were proposed in \cite{Kouvaris:2014uoa}.

Our paper can be viewed as a significant complement to earlier literature on singlet fermion DM \cite{LopezHonorez:2012kv, Fairbairn:2013uta,Alves:2013tqa,Alanne:2014bra,Kouvaris:2014uoa}: We investigate the effects of both scalar and pseudoscalar couplings between the DM fermion and the scalar mediator. Via Monte Carlo scan, we probe wide mass scales for the singlet sector especially focusing on the range where the DM self-interaction cross section is large enough to reconcile the observations and simulations of small scale structure. In addition, we take systematically into account the phenomenological constraints from DM abundance and direct detection as well as theoretical constraints of stability and perturbativity. Furthermore, we confirm that the BBN constraint can be resolved in this model via the mechanism proposed in \cite{Kouvaris:2014uoa}.

The paper is organized as follows: In section \ref{model} we define the model and outline the basic computations constraints on DM abundance and direct detection. In section \ref{selfinteraction} we study the strength of DM self-interactions and its relation to the mass of the scalar mediator. The consequences of the existence of a light mediator on the physics of the early Universe is discussed in section \ref{cosmo}. In section \ref{checkout} we present our conclusions and outlook.

%
\section{Model}
\label{model}
%
\subsection{Scalar sector and the dark matter Lagrangian}

We consider an extended scalar sector described by the potential
\bea
V(\phi,s)=&\mu_\phi^2\phi^\dagger \phi+\lambda_\phi(\phi^\dagger \phi)^2+\mu_1 s +\frac{\mu_s^2}{2} s^2+\frac{\mu_3}{3} s^3+\frac{\lambda_s}{4}s^4\\
&+\muphis (\phi^\dagger \phi)s+\frac{\lphis}{2}(\phi^\dagger \phi)s^2,
\label{scalarpotII}
\eea
which provides typical scalar portals between the hidden singlet sector and the SM Higgs. The field $\phi$ is the usual SM Higgs doublet and $s$ is a real singlet scalar. The Higgs doublet is written in terms of its components as
\be
\phi=\begin{pmatrix} \chi^+\\ \frac{1}{\sqrt{2}}(v+h+\mathrm{i} \chi)\end{pmatrix},
\ee
where $v$ is the vacuum expectation value which at $T=0$ has the value $v = 246\GeV$. The stability of the tree level potential requires that 
\be
\lambda_\phi>0, \quad \lambda_s>0, \quad \lphis>-2\sqrt{\lambda_\phi \lambda_s}.
\ee
To further constrain the parameters we also require that the electroweak broken minimum is the deepest one at $T = 0$.  There are several inequivalent minima whose relative ordering can evolve as a function of temperature. As shown in \cite{Alanne:2014bra} this potential leads to strong first order electroweak transition over a large portion of the parameter space already at the simple mean field level. 

In this paper we do not consider the electroweak transition in detail, but focus on the properties of the DM candidate, which we take to be a singlet fermion described by the Lagrangian
\be
{\cal L}_{\rm{DM}}=\bar{\psi}(\mathrm{i}\slashed\partial-m_\psi)\psi+s\bar{\psi}\left(g_s + \i g_p \gamma_5\right)\psi.
\label{flagr}
\ee
Without loss of generality, we can shift $s\rightarrow s^\prime=s-\langle s\rangle$ such that at the electroweak broken minimum $\left\langle s^\prime \right\rangle=0$. Since we are not assuming any discrete symmetries for the singlet scalar, the shift will simply reproduce all terms already present in the scalar potential with redefined coefficients. Hence practically we use (\ref{scalarpotII}) and (\ref{flagr}), where the mass of the DM fermion at $T<T_\mathrm{EW}$ is given directly by the mass parameter $m_\psi$ and the mixing between $\phi$ and $s$ fields is determined solely by $\muphis$. In principle $m_\psi$ is a complex number, but we can always make it real by a chiral rotation on $\psi$ in Eq. (\ref{flagr}).\footnote{Alternatively, we could rotate away either of the couplings $g_{s,p}$, but then we would need to allow for complex $m_\psi$.}

We use the extremization conditions to set
\be
\mu_\phi^2 = - v^2 \lambda_\phi \,, \quad \mu_1 = -\frac{v^2 \muphis}{2}.
\ee
Diagonalizing the mass matrix 
\be
M^2 = \left(
\begin{array}{cc}
2 v^2 \lambda_\phi  & v \muphis \\
v \muphis & \frac{v^2 \lphis}{2}+\mu_s^2 \\
\end{array}
\right),
\ee
we trade the parameters $\lambda_\phi$ and $\mu_s^2$ with physical masses $m_H = 126$ GeV and $m_S$,
\bea
\label{trade}
&\lambda_\phi = \frac{1}{4 v^2}\left(m_H^2+m_S^2\pm\sqrt{(m_H^2-m_S^2)^2-4 \muphis^2 v^2}\right), \\
&\mu_s^2 = \frac{1}{2} \left(m_H^2+m_S^2-\lphis v^2\mp\sqrt{(m_H^2-m_S^2)^2-4 \muphis^2 v^2}\right).
\eea
The scalar mass eigenstates at $T=0$ are 
\be
H = h \cos\beta + s \sin\beta, \quad S = -h \sin\beta + s \cos\beta,
\label{scalarmixing}
\ee
where $\tan 2\beta = -4v\muphis/(2\mu_s^2 + \lphis v^2 - 4\lambda_\phi v^2)$. Hence, the free parameters are $m_{\psi}$, $m_S$, $\mu_3$, $\muphis$, $\lambda_s$, $\lphis$, $g_s$ and $g_p$. From equation \eqref{trade} it follows that $|\muphis| < \max(m_H,m_S)^2/2 v$.

%
\subsection{Constraints}
%

The Higgs coupling data from the LHC and Tevatron experiments constrains the mixing angle $\cos\beta$ and the decay width of the Higgs boson. The invisible decay width of the Higgs is given by
\be
\Gamma_{\rm{inv}} = \frac{\lambda_\mathit{HSS}^2v_S}{32\pi m_H} \Theta(m_H-2m_S)
+ \frac{m_H\sin^2\beta}{4\pi} \Big( g_s^2 v_\psi^3 + g_p^2 v_\psi \Big) \Theta(m_H-2m_\psi) ,
\ee
where $v_a^2 \equiv 1-4m_a^2/m_H^2$ and the coupling $\lambda_\mathit{HSS}$ is given in the appendix \ref{xsect}. We perform a $\chi^2$ fit to the Higgs boson signal strength data from the LHC and Tevatron experiments \cite{CMS-PAS-HIG-14-009, Aad:2014eva, ATLAS:2014aga, Aad:2014xzb, ATLAS-COM-CONF-2014-080, Aad:2014eha, Aaltonen:2013kxa} and require that the model remains within $2\sigma$ of the best fit. Moreover we check that the model is compatible with the electroweak precision measurements using the $S$ and $T$ parameters \cite{Peskin:1990zt,Agashe:2014kda}. 

We also study the perturbativity of the model as a function of the energy scale. To be concrete, we require that all models remain perturbative up to $10$ TeV, but we find that many models remain perturbative up to Planck scale. The 1-loop $\beta$ functions for the Yukawa couplings, $g_s$ and $g_p$, are
\begin{align}
16\pi^2\beta_{g_s}=&5g_s^3+3g_p^2g_s,\\
16\pi^2\beta_{g_p}=&3g_p^3+5g_s^2g_p,
\end{align}
whereas the $\beta$ functions of the quartic couplings read
\begin{align}
\begin{split}
16\pi^2 \beta_{\lambda_\phi}=& 24\lambda_\phi^2+\frac{1}{2}\lphis^2-3\left(3g_{\mathrm{L}}^2+g_Y^2-4y_t^2\right)\lambda_\phi\\
&+\frac{3}{8}\left(3g_{\mathrm{L}}^4+2g_{\mathrm{L}}^2g_Y^2+g_Y^4\right)-6y_t^4, 
\end{split} \\
\begin{split}
16\pi^2\beta_{\lphis}=& 4\lphis^2+\left(12\lambda_\phi+6\lambda_s\right)\lphis\\
&-3\left(\frac{3}{2}g_{\mathrm{L}}^2+\frac{1}{2}g_Y^2-2y_t^2-\frac{4}{3}g_s^2\right)\lphis,
\end{split} \\
\begin{split}
16\pi^2\beta_{\lambda_s}=&18\lambda_s^2+2\lphis^2+8\lambda_s g_s^2-6 \left(g_s^2+g_p^2\right)^2.
\end{split}
\end{align}
The $\beta$ functions for the top-coupling $y_t$ and the weak and hypercharge gauge couplings $g_L$ and $g_Y$ are standard, see for example ref.~\cite{Shaposhnikov:2009pv}.

Although one can study if the model remains perturbative and maintains vacuum stability all the way to the Planck scale, the question is not obviously meaningful as the approximation of neglecting curvature effects may not be valid \cite{Herranen:2014cua}. Hence, we find it more natural to treat this dark sector as a low energy realization of some more complete theory. The relevant matching scale is of course completely unknown at this stage; a possible concrete UV completion is for example the model introduced in \cite{Kainulainen:2015raa}. 

Finally, direct DM searches give an upper limit on DM scattering off nuclei. Since $\psi$ interacts with SM only through the Higgs exchange, the relevant detection channel is the spin-independent (SI) one. The nominal SI cross section in this model is given by:
\be
\sigma_{\rm{SI}}^0 = \frac{\mu_N^2 f_N^2 m_N^2}{\pi v^2} g_s^2 \, {\sin}^2\beta\, {\cos}^2\beta \left(\frac{1}{m_H^2} - \frac{1}{m_S^2} \right)^2,
\label{eq:SigmaSI}
\ee
where $m_N$ is the nucleon mass, $\mu_N$ is the reduced mass of the WIMP-nucleon system and the Higgs-nucleon coupling $f_N \approx 0.303$. However, we do not require that $\psi$ forms {\em all} of the existing DM. We therefore define the relative DM relic density, $f_\mathrm{rel} = \Omega h^2/0.12$. Consequently, we define an effective cross section $\sigma_{\rm SI}^{\rm eff} \equiv f_{\rm rel} \sigma_{\rm SI}^0$, which is the quantity directly constrained by observations for a subleading DM component~\cite{Cline:2012hg}. Currently the most restrictive upper limits come from the LUX and SuperCDMS experiments~\cite{Akerib:2013tjd,Agnese:2014aze}. 

%
\subsection{Freeze-out}
%

To evaluate the relative DM abundance $f_\mathrm{rel}$, we concentrate to the region of parameters where $\psi$ is a thermal relic that decouples with the freeze-out mechanism~\footnote{Another possibility, requiring much smaller couplings $g_{p,s}$ than the ones considered here is that $\psi$ was ever only partially thermalized by the freeze-in mechanism.}. The annihilation processes are $\psi\overline\psi \to HH,SS,HS$ and $\psi\overline\psi \to XY$, where $X,Y$ are SM fermions or gauge bosons. The latter are $s$ channel processes with $H$ or $S$ propagators. We calculate the cross section in terms of the off-shell Higgs boson decay width:
\be
v_{\psi s} \sigma_{\psi\overline\psi \to XY} =  \frac{1}{2}{\sin}^2\beta {\cos}^2\beta 
\Big(g_s^2 v_{\psi s}^2 + g_p^2 \Big) \left| D_H - D_S \right|^2 \sqrt{s} \,\Gamma_h(\sqrt{s}),
\ee
where $v_{\psi s}^2 \equiv 1-4m_\psi^2/s$ and $D_a^{-1} \equiv s-m_a^2 + \mathrm{i} m_a \Gamma_a$.
We evaluate $\Gamma_h(\sqrt{s})$ following~\cite{Cline:2013gha}: for $\sqrt{s}<$ 80 GeV and $\sqrt{s}>$ 300 GeV we use perturbative decay widths taken from~\cite{hunters} including the QCD corrections for the quark final states. In the range $\sqrt{s}\in(80\GeV,300\GeV)$ we use the tabulated results from \cite{Dittmaier:2011ti}, which include also four body final states. As shown in \cite{Clarke:2013aya} there are large uncertainties in the range $\sqrt{s} \in (2m_\pi,2m_c)$. In this region we use the perturbative spectator approach \cite{hunters}. The cross section for the scalar final states is given in appendix \ref{xsect}. 

Note that the annihilation cross section contains contributions both from the scalar and pseudoscalar couplings $g_s$ and $g_p$, while the cross section for DM scattering on a nucleus is proportional only to $g_s$. This allows one to escape the direct detection limits by reducing $g_s$ while maintaining the large enough annihilation cross section by adjusting the value of $g_p$.

Given the cross sections, we solve the thermal relic abundance, $Y\equiv n_\psi/s$ (where $s$ is the entropy density) from the Zeldovish-Okun-Pikelner-Lee-Weinberg equation \cite{Zeldovich:1965,Lee:1977ua}
\be
\frac{\mathrm{d}Y}{\mathrm{d}x} = Z \left(Y^2 - Y_\mathrm{eq}^2\right)\;.
\ee
Here $x \equiv m_\psi/T$ and
\be
Z(x) = -\sqrt{\frac{\pi}{45} g_*(T)} \frac{M_\mathrm{Pl} m_\psi}{x^2} \left\langle v \sigma  \right\rangle \,,
\ee
where
\be
g^{1/2}_*(T) = \frac{h_\mathrm{eff}}{\sqrt{g_\mathrm{eff}}} \left( 1+\frac{T}{3h_\mathrm{eff}} \frac{\mathrm{d}h_\mathrm{eff}}{\mathrm{d}T} \right)
\ee
and $g_\mathrm{eff}$ and $h_\mathrm{eff}$ are the energy and entropy degrees of freedom, respectively. For the averaged cross sections, we use the integral expression \cite{Gondolo:1990dk}
\be
 \left\langle v \sigma \right\rangle = \frac{1}{8 m_\psi^4 T K_2^2\left(x\right)} \int_{4m_\psi^2}^\infty \mathrm{d}s \sqrt{s} (s-4m_\psi^2) K_1\left(\frac{\sqrt{s}}{T}\right) \sigma_\mathrm{tot}(s)\;,
\ee
where $K_i(y)$ are the modified Bessel functions of the second kind. Finally the relative DM relic density, $f_\mathrm{rel} = \Omega h^2/0.12$, is given by 
\be
f_\mathrm{rel} = 2.30\cdot10^9 \frac{m_\psi}{\mathrm{GeV}} Y(x_0),
\ee
where $x_0 = m_\psi/T_0$ and $T_0$ is the photon temperature today.

%
\section{Dark matter self-interaction}
\label{selfinteraction}
%

The DM self-interaction processes are $\psi\overline\psi\to\psi\overline\psi$, $\overline\psi\overline\psi\to\overline\psi\overline\psi$ and $\psi\psi\to\psi\psi$. The first of these arises from the $s$ and the $t$ channel, and the latter two from $t$ and $u$ channel diagrams with $H$ and $S$ propagators. It turns out that to obtain reasonably strong DM self-interaction $S$ has to be very light, $m_S \lesssim 1\GeV$, and the contribution of diagrams with $H$ propagator  can be neglected. In the interesting region the Born approximation for the self-interaction fails and we must calculate the cross section more accurately by solving the Schr\"{o}dinger equation for Yukawa potential,
\be
V(r) = -\frac{\alpha}{r} e^{-m_S r},
\ee 
where $\alpha = {\cos}^2\beta \, g_s^2/4\pi$. In the partial wave analysis the scattering amplitude written in terms of phase shifts $\delta_l$ is given by
\be
f(\theta) = \frac{1}{k} \sum_{l=0}^\infty (2l+1) e^{\i \delta_l} P_l(\cos\theta) \sin\delta_l,
\ee
where $k = \mu v = m_\psi v/2$, $v$ is the relative velocity and $P_l$ is the $l$th Legendre polynomial. To find the phase shifts we need to solve the radial Schr\"{o}dinger equation,
\be
\frac{1}{r^2} \frac{\mathrm{d}}{\mathrm{d} r} \left( r^2 \frac{\mathrm{d} R_l}{\mathrm{d} r} \right) + \left( k^2 - \frac{l(l+1)}{r^2} - 2 \mu V(r) \right) R_l = 0,
\ee
and match the solution onto the asymptotic solution,
\be
\lim_{r\to\infty}R_l \propto \cos\delta_l \, j_l(k r) - \sin\delta_l \, n_l(k r),
\ee
where $j_l$ and $n_l$ are the spherical Bessel and spherical Neumann functions, respectively. In the numerical solution we use the method described in \cite{Tulin:2013teo}. 

Combining all self-interaction processes the total differential cross section is finally given by
\be
\frac{\mathrm{d}\sigma_\psi}{\mathrm{d}\Omega} = 2 \left| f(\theta) - f(\pi-\theta) \right|^2 + \left| f(\theta) \right|^2 \,,
\ee
where the first term comes from the $\psi\psi$- and $\overline\psi\overline\psi$-channels, which need to be antisymmetrized over the scattering angle $\theta$, and the last term represents the $\overline\psi\psi$-channel.

\begin{figure}
\begin{center}
\includegraphics[width=0.6\textwidth]{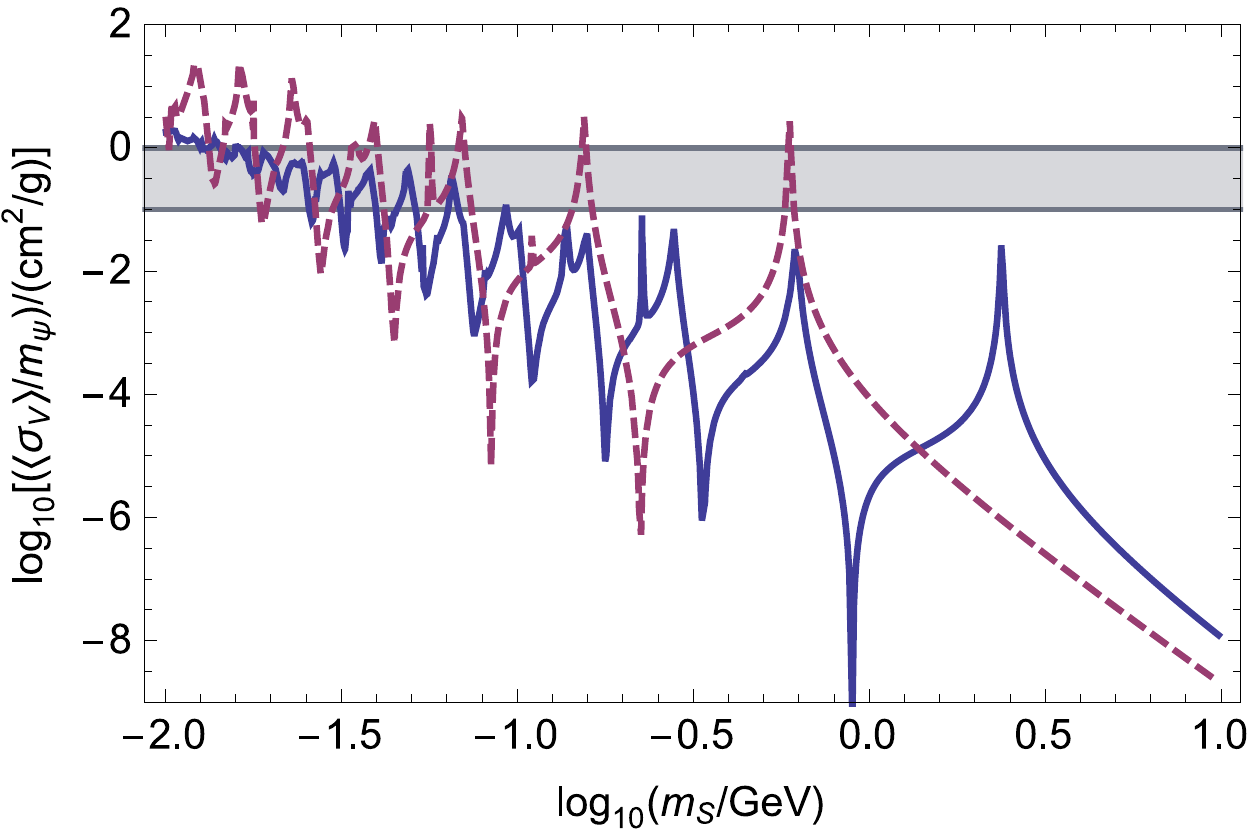}
\caption{Self-interaction as a function of $m_S$ for fixed $\alpha = 0.01$ and $\Delta v = 10\,\mathrm{km}/\mathrm{sec}$. Solid and dashed lines correspond to $m_\psi = 400\GeV$ and $m_\psi = 100\GeV$, respectively. The shaded band corresponds to $\left\langle \sigma_V \right\rangle/m_\psi \sim 0.1-1 \, \mathrm{cm}^2/\mathrm{g}$.}
\label{siplot}
\end{center}
\end{figure}

The self-interactions of DM particles are essential input for numerical simulations of DM halo formation. The scattering angle dependence of the differential cross section is particularly important when the particle trajectories are tracked through collisions. Indeed, while the integrated cross section is strongly enhanced in the forward region for light mediators, the forward collisions are unimportant as they leave the particle trajectories unaffected. Hence, to parametrize transport properties one uses instead the viscosity cross section
\be
\sigma_V = \int \mathrm{d}\Omega \sin^2\theta \frac{\mathrm{d}\sigma_\psi}{\mathrm{d}\Omega} \,,
\ee
which in general is some function of the relative velocity of the DM particles. Moreover, we average the viscosity cross section over the thermal DM velocity distribution:
\be
\left\langle \sigma_V \right\rangle = \int \frac{\mathrm{d}^3 v}{(4\pi \Delta v^2)^\frac{3}{2}} \,e^{-v^2/4\Delta v^2} \,\sigma_V,
\ee
where $\Delta v$ is the radial velocity dispersion of the halo. We use $\Delta v = 10\, \mathrm{km}/\mathrm{sec}$, corresponding to observations of stellar motions in local dwarf galaxies~\cite{Walker:2007ju}. In figure \ref{siplot} we show the dependence of the self-interaction cross section on $m_S$ for fixed parameters. In order to be consistent with astrophysical observations, cosmological simulations require $\left\langle \sigma_V \right\rangle/m_\psi \sim 0.1-1 \,\mathrm{cm}^2/\mathrm{g}$ \cite{Rocha:2012jg,Peter:2012jh,Zavala:2012us}, and this region is shown as a shaded band in figure \ref{siplot}.

%
%

If $m_S<m_H/2$ the LHC limits on the invisible Higgs decays constrain the parameter space significantly. Also the spin-independent DM-nucleon cross section, $\sigma_\mathrm{SI}$, which is proportional to $1/m_S^4$, gives a strong constraint. For a very light $S$ the latter is by far the most stringent of the two. Avoiding these limits requires a very small effective coupling $g_s \sin\beta$ when $S$ is very light. However, $g_s$ itself should not be small, because we want to have a large self-interaction rate. It follows that the mixing angle $\beta$ has to be very small.

In order to study the feasibility of the model, we performed Monte Carlo scan of the parameter space with the following priors:
\be
|g_s|<1, \qquad |g_p|<1, \quad 0<\lambda_s<\pi, \quad -9<\log|\lphis|<0,  \\ 
\ee
and
\be
\begin{gathered}
-9 < \log|\muphis/\GeV|<1.5\,, \qquad -5 < \log|\mu_3/\GeV| < 0\phantom{.} \\ 
1\GeV<m_{\psi}<1000\GeV,\qquad  -2<\log(m_S/\GeV)<0.
\end{gathered}
\ee

\begin{figure}
\begin{center}
\includegraphics[width=0.45\textwidth]{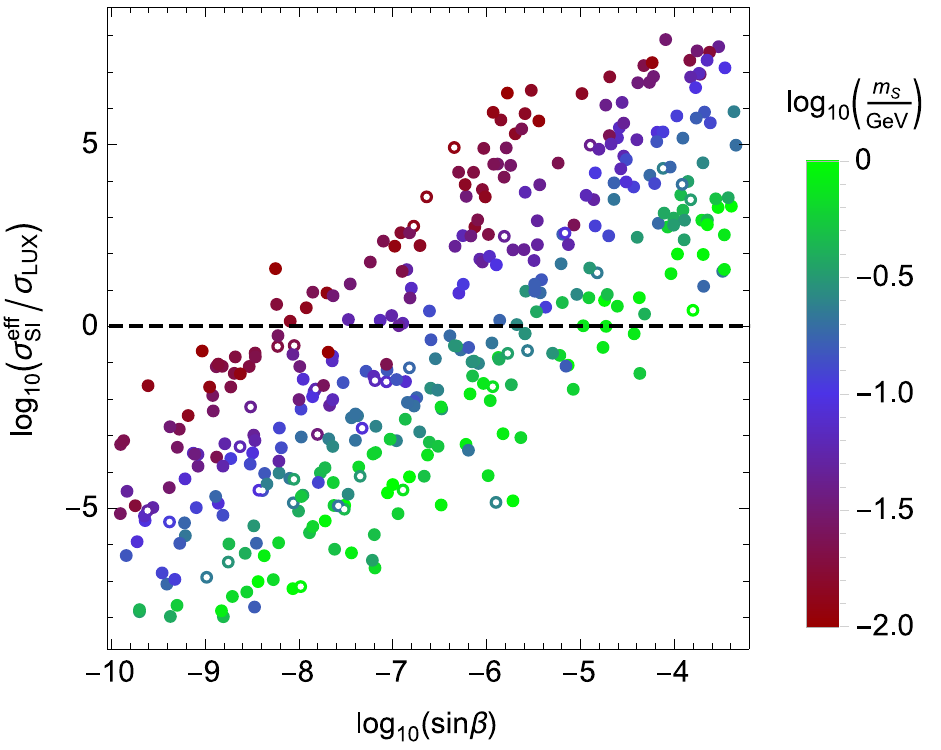} \hspace{0.2cm}
\includegraphics[width=0.45\textwidth]{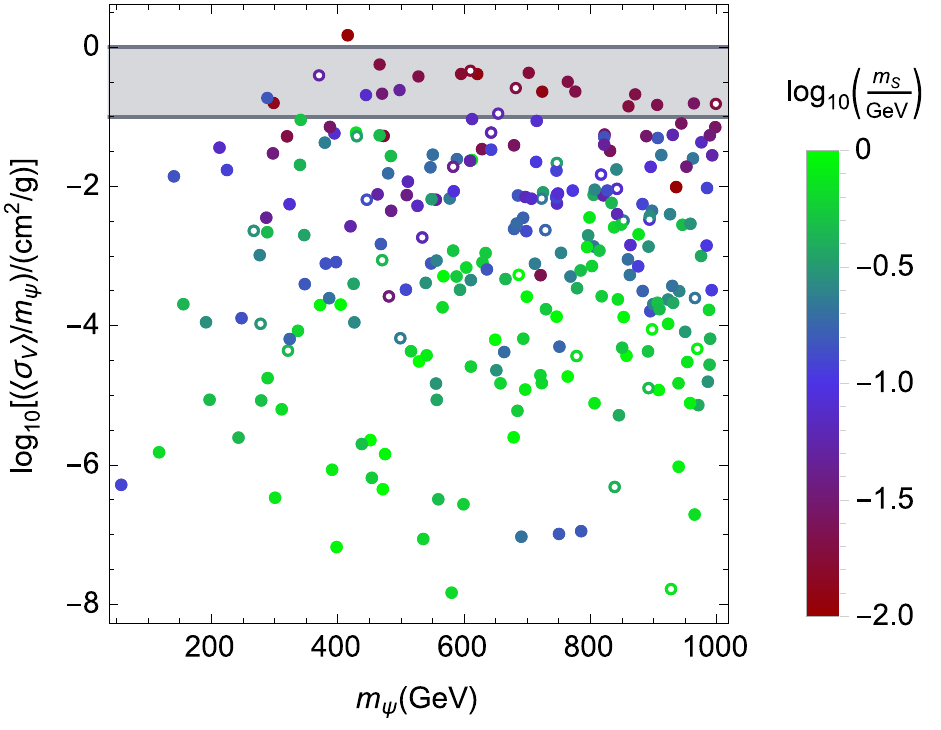}
\caption{Shown are the parameter sets which are in agreement with collider constraints and give a relic density in the range $0.8 < f_\mathrm{rel} \leq 1$. In the right panel all points are below the LUX limit. Unfilled points indicate models that remain perturbative up to Planck scale.}
\label{foresults}
\end{center}
\end{figure}

The resulting parameter sets, which are in agreement with the collider constraints are shown in figure \ref{foresults} as a function of the mixing angle $\sin \beta$ and the SI detection cross section normalized to LUX sensitivity (left) and as a function of the DM particle mass $m_\psi$ and self-interaction cross section (right). The color coding of dots specifies the mediator mass $m_S$ as indicated by the vertical bar to the right of the panels. To be consistent with the LUX limit~\cite{Akerib:2013tjd} the mixing angle has to roughly  
satisfy a constraint 
\be
\sin\beta\lesssim10^{-5} (m_S/\mathrm{GeV})^2 \,,
\label{eq:roughconst}
\ee
consistent with Eq.~(\ref{eq:SigmaSI}).

%
%

%
\section{Cosmology of a light long-lived mediator}
\label{cosmo}
%

We found that in the region interesting for DM self-interactions the mediator is light, $m_S\lesssim 0.1\GeV$, and its coupling to SM fermions is very small, $\sin\beta \lesssim 10^{-5}$. Since $S$ decay to SM particles is suppressed by the small mixing angle, its lifetime tends to be very long and it may cause problems in the early Universe \cite{Kaplinghat:2013yxa,Zhang:2015era}. In particular there will be an upper bound on the $S$-lifetime from nucleosynthesis.

%
\subsection{Bound from the Big Bang nucleosynthesis}
%

We will first discuss the BBN bound in the case where $s$ decays only to SM particles.  
Here $S$ decays predominantly to the heaviest available SM fermion with a width:
\be
\Gamma_{S\to f\overline{f}} = \frac{m_f^2 \,{\sin}^2\beta }{8\pi v^2} \,m_S\left( 1-\frac{4m_f^2}{m_S^2} \right)^{3/2} \,,
\ee
where $v \approx 246$ GeV and $m_f$ is the  mass of the fermion. This corresponds to a lifetime

\be
\tau_{S} \; \approx \; \left(\frac{10^{-6}}{{\sin}\beta}\right)^2 
                                  \left(\frac{\rm MeV}{m_f}\right)^2
                                  \left(\frac{\rm GeV}{m_S}\right) \rm sec\,.
\label{eq:Slifetime}
\ee
For $m_S \gsim 2m_\mu$, $S$ decays to muons and for $2m_\mu > m_S > 2m_e$ the dominant channel is to electrons and positrons. 

When $S$ decays to SM particles, the decay products in general thermalize quickly. An exception may be neutrinos, which decouple from equilibrium at $T\approx 4.3$ MeV (muon- and tau neutrinos) and $T \approx 2.6$ MeV (electron neutrinos)~\cite{Enqvist:1991gx}. Thus, if decays take place around 2.3 MeV, corresponding to $\tau_S \gsim 0.1$ sec, they may lead to an increase of the electron neutrino density~\cite{Fields:1995mz}. This would increase the rate of reactions $\nu_e + n \leftrightarrow p + e^-$ and $\bar\nu_e + p \leftrightarrow n + e^+$, which keep protons and neutrons in thermal equilibrium and hence decrease the $\mbox{}^4$He abundance~\cite{Enqvist:1991qj}. For longer lifetimes this neutrino heating effect becomes rapidly weaker. Instead, for $\tau_S \gsim 1$ sec, the residual population of $S$-particles starts to increase the expansion rate of the Universe. This causes the proton-neutron equilibrium reactions to drop out of equilibrium earlier, which makes the $\mbox{}^4$He abundance larger. A fully quantitative analysis of these effects is beyond the scope of this paper. Instead we follow~\cite{Kawasaki:2000en} and, to remain within the $2\sigma$ limit of the observed $\mbox{}^4$He abundance, require that $\tau_S < 1 \,\mathrm{sec}$. 

\begin{figure}
\begin{center}
\includegraphics[width=0.45\textwidth]{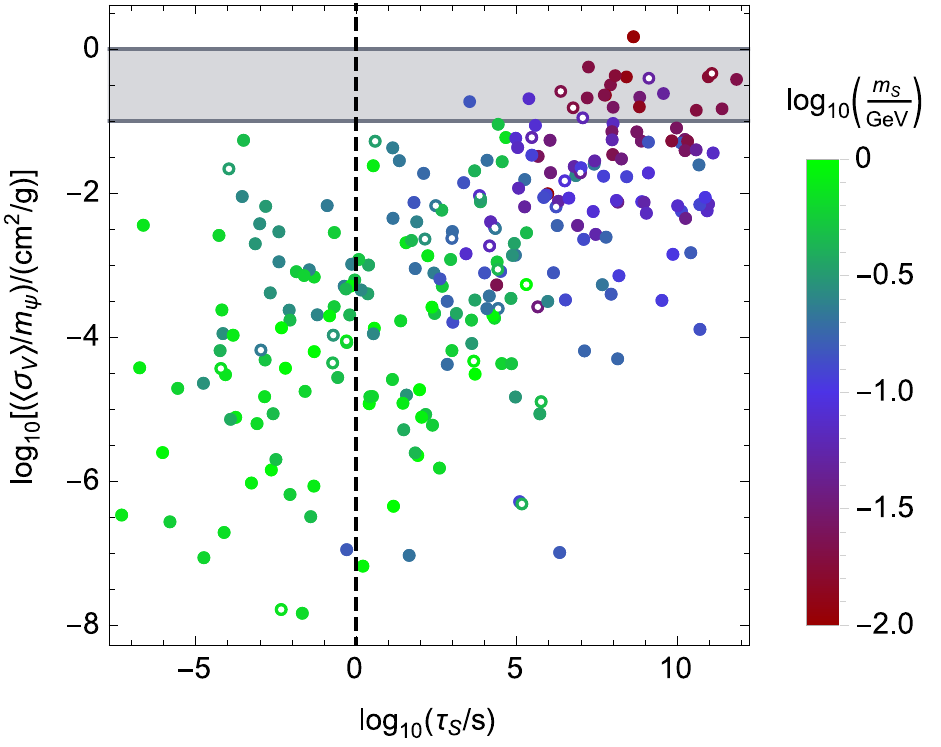}
\includegraphics[width=0.45\textwidth]{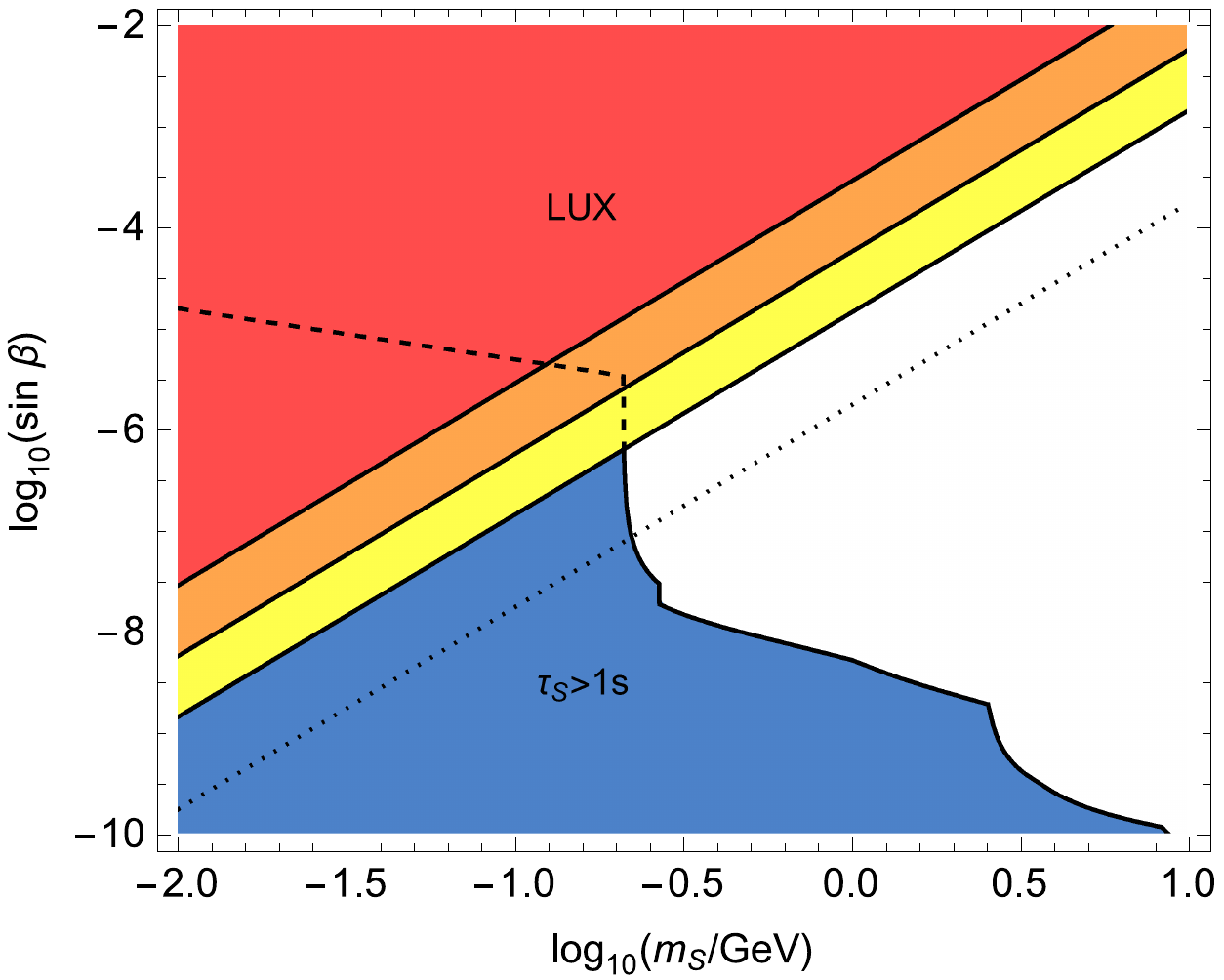}
\caption{Left: Shown are the models which are in agreement with collider constraints and the LUX limit and give $0.8 < f_\mathrm{rel} \leq 1$ as a function of the $S$-lifetime and the self-interaction cross section $\left\langle \sigma_V \right\rangle/m_\psi$.  The vertical dotted line shows the $\tau_S = 1\,\mathrm{sec}$ limit and the shaded region corresponds to models with sufficient self-interactions. Unfilled points again indicate models that remain perturbative up to Planck scale. Right: The excluded regions in the mixing angle versus the scalar mass from BBN (blue area) and from the LUX constraints (yellow, orange, red) corresponding to $g_s=0.4$, $0.1$, $0.02$, respectively. The DM mass is fixed to $400\GeV$. The dotted line indicates the expected XENON1T experiment reach~\cite{Aprile:2012zx} in the case $g_s=0.4$.}
\label{BBNLUXexcl}
\end{center}
\end{figure}

Imposing the limit (\ref{eq:roughconst}) on the mixing angle, equation (\ref{eq:Slifetime}) gives a rough bound on the $S$-particle lifetime: 
\be
\tau_S \gtrsim 0.01 \, \left(\frac{\rm MeV}{m_f}\right)^2\left(\frac{\rm GeV}{m_S}\right)^5 \;\rm sec \,.
\ee 
Since for $m_S \lesssim 0.2\GeV$ the dominant decay channel is to electrons, it is clear that a light mediator with $m_S \lesssim 0.1\GeV$, decaying to SM particles, is not consistent with the BBN and direct search constraints. This conclusion becomes evident also from figure \ref{BBNLUXexcl}, where the bounds on the mixing angle and the mediator mass following from the BBN and LUX constraints are shown more accurately.

%
%


\subsection{Extension with a light sterile neutrino}

An obvious solution to alleviate this problem is to introduce new states into which a light $S$ can decay to, such as a light sterile neutrino $N$ which couples to the singlet scalar via a Lagrangian~\cite{Kouvaris:2014uoa}, 
\be
\mathcal{L}_{\rm SNN} = y_N SN\bar{N} \,.
\ee
If $m_N < m_S/2$ and $N$ mixes with the SM neutrinos with a small, but finite mixing angle $\sin\theta$, it is possible to extend the acceptable parameter space. Let us now consider 
in quantitative detail how this mechanism works in this model.

\begin{figure}[tb]
\begin{center}
\includegraphics[width=0.6\textwidth]{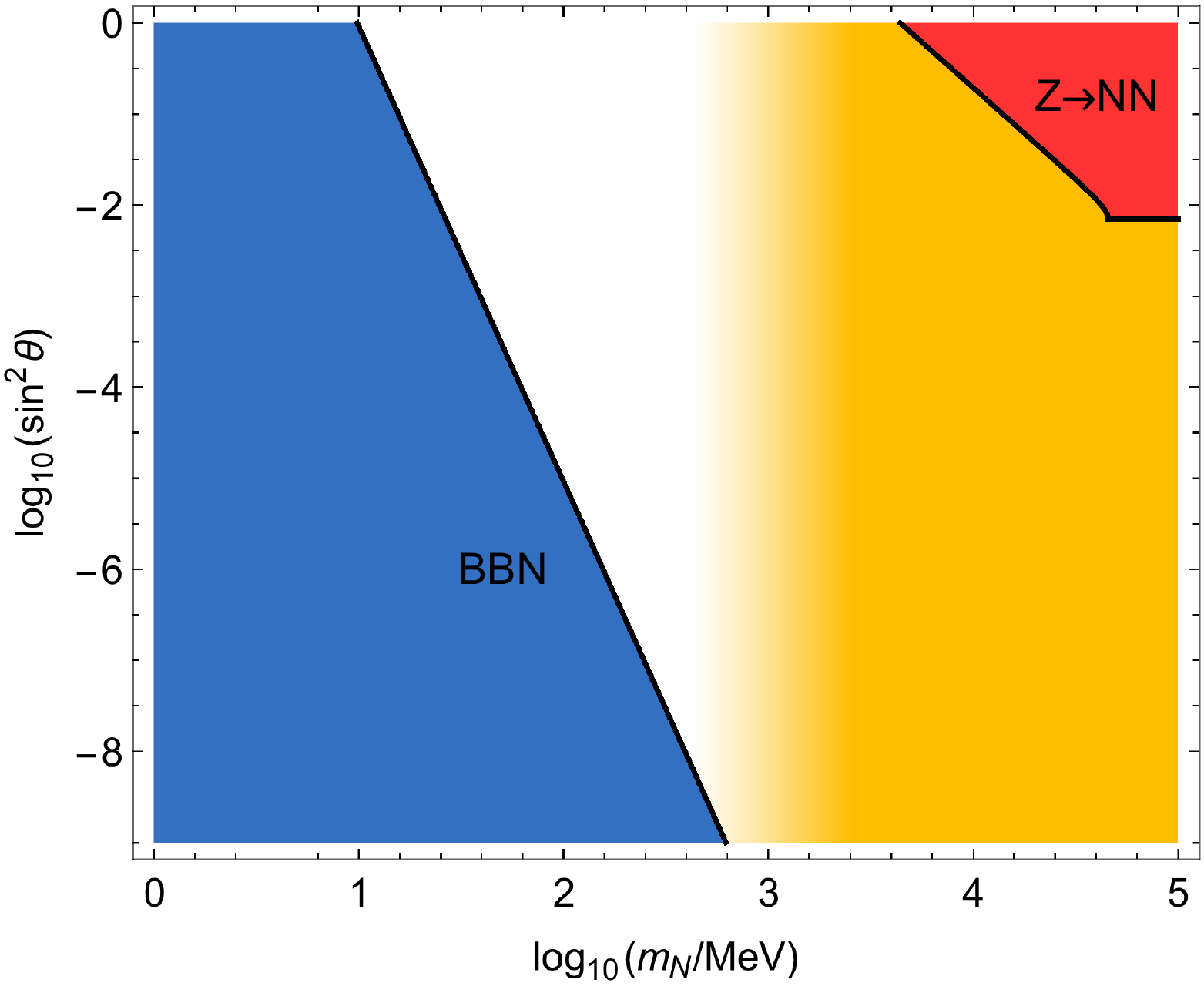}
\caption{Shown are the excluded regions in the sterile $N$ parameters due to the $Z$-decay width (red area) and the BBN (blue area). In the
rightmost shaded region (yellow area) $S$ is too heavy to provide a sufficiently strong self-interaction for $\psi$.}
\label{sterNplot}
\end{center}
\end{figure}

First, an upper bound for the mixing $\sin\theta$ is obtained from the invisible decay width of the $Z$ boson, experimentally determined to be \cite{Agashe:2014kda}
\be
\frac{\Gamma(Z\to\mathrm{inv})}{\Gamma(Z\to\nu\nu)} = 2.990\pm 0.007.
\ee
The invisible $Z$ boson decay width is
\be
\Gamma(Z\to{\rm inv}) = \Gamma(Z\to\nu_e\nu_e) + \Gamma(Z\to\nu_\tau\nu_\tau) + \Gamma(Z\to\nu_\mu\nu_\mu) + \Gamma(Z\to NN).
\ee
If for example $N$ mixes only with $\tau$ neutrino, then
\bea
&\Gamma(Z\to\nu_e\nu_e) = \Gamma(Z\to\nu_\mu\nu_\mu) = \Gamma(Z\to\nu\nu), \\&\Gamma(Z\to\nu_\tau\nu_\tau) = \cos^2\theta\, \Gamma(Z\to\nu\nu), \\
&\Gamma(Z\to NN) = \sin^2\theta\, \Gamma(Z\to\nu\nu) \left( 1-\frac{m_N^2}{m_Z^2} \right) \left( 1-\frac{4m_N^2}{m_Z^2} \right)^{\frac{1}{2}}.
\eea
Now for $m_N\ll m_Z$ the invisible decay with of $Z$ is $\Gamma(Z\to{\rm inv}) = 3 \Gamma(Z\to\nu\nu)$, as in the SM. This decay width is in agreement with the experimental value. The mixing modifies $\Gamma(Z\to{\rm inv})$ only at $m_N\gsim m_Z$, as shown by the red area in figure~\ref{sterNplot}.

Second, we require that the $S$ decays to sterile neutrinos sufficiently fast and also that the sterile neutrinos have time to decay to SM neutrinos before they freeze-out.  To be precise, we first require that the lifetimes of both $S$ and $N$ are less than $0.1 \,\mathrm{sec}$. The decay width for the process $S\to N\bar{N}$ is the same as in the SM fermion case:
\be
\Gamma_{S\to N\bar{N}} = \frac{y_N^2 }{16\pi} m_S \left( 1-\frac{4m_N^2}{m_S^2} \right)^{3/2},
\ee
so the lifetime of $S$ is smaller than $0.1 \, \mathrm{sec}$ if 
\be
y_N 
\gsim 2\times 10^{-11} \left( \frac{\GeV}{m_S} \right)^{1/2} \,.
\ee
Clearly $y_N$ can arranged such that $S$ decays fast enough. We have checked explicitly that the new annihilation channel induced by this coupling does not affect the DM abundance. On the other hand, the main decay channel for the sterile neutrino is to three SM neutrinos. The decay width for this process is
\be
\Gamma_{N\to 3\nu} = \frac{G_F^2}{192\pi^2} m_N^5 \sin^2\theta \,.
\ee
The constraint that the lifetime of $N$ is smaller than $0.1 \; \mathrm{sec}$ is then expressed as
\be
m_N (\sin \theta)^{2/5} > 10\MeV\,. 
\label{eq:BBN-bound}
\ee
Hence adding a new sterile neutrino $N$ and requiring it to decay to SM neutrinos before they drop out of thermal equilibrium relieves the BBN limit for $m_S > 20\MeV$. Nominally, this limit excludes the region to the left from the rightmost solid line in figure~\ref{sterNplot}.

%
%

Since $N^\prime$s decay only to neutrinos, their late decay could lead to a significant increase in the neutrino energy density, potentially in conflict with the existing bounds on the amount of dark radiation~\cite{Rossi:2014nea}. We will therefore require that $N$ energy density does not exceed that of one half equivalent neutrino degree of freedom at the time of their decay. Assuming that the freeze-out of sterile neutrinos occurred before the QCD phase transition ($T_{\rm fo} >T_{\rm QCD}$), we find that at $T < m_e$ 
\be
 \rho_N = m_N n_N = m_N \frac{g_{\rm eff}(T)}{g_{\rm eff}(T_{\rm fo})} \frac{3\zeta(3)}{2\pi^2} T^3 \approx 0.008 m_N T^3,
 \ee
which relative to the relativistic neutrinos is
 \be
 \frac{\rho_N}{\rho_\nu} = \frac{0.008m_NT^3}{\frac{7\pi^2}{120}\left(\frac{4}{11}\right)^{4/3} T^4} \approx 0.05 \frac{m_N}{T}.
 \ee
We then must require that $N$ decays at the latest at $T = m_N/10$.  The $N$ decay decay width relative to the Hubble expansion rate is given by
\be
 \frac{\Gamma_{N\to3\nu}}{H}\Big|_{10T=m_N} \approx 0.03 \left(\frac{m_N}{\rm MeV}\right)^3 \sin^2\theta,
\ee
which gives
\be
 m_N\left(\sin\theta\right)^{2/3} > 3 \;{\rm MeV}.
\ee
Hence, the ultimate lower bound for the mass of $N$ at given $\sin^2\theta$ comes from the requirement that $N$ decays before BBN.
 
%
\section{Conclusions}
\label{checkout}
%

In this paper we have considered a model for self-interacting DM. The DM candidate was taken to be a fermion $\psi$ with self-interactions mediated by a scalar $S$. The scalar also mediates coupling between the hidden and visible sectors via the usual Higgs portal. We demonstrated that over a large region of parameter space the model can saturate the observed DM relic density. For the self-interactions to be sufficiently strong, the scalar mediator has to be light which leads to the well-known problems in the early Universe if the scalar survives until the era of BBN. We showed that the pseudoscalar coupling $g_p$ makes it easier to get the correct DM abundance and to simultaneously escape the direct detection limits by reducing $g_s$, but since the DM self-interaction cross section is dominantly proportional only to $g_s$, this does not alleviate the problem with DM direct detection and BBN limits in getting sufficiently strong DM self-interaction. We confirmed the observation \cite{Kouvaris:2014uoa} that this problem can be solved by assuming the existence of a light sterile neutrino coupling with the singlet scalar and mixing with the active neutrinos. The allowed region for the extended model consists of a triangle at 
$10 (\sin \theta)^{-2/5} \,{\rm MeV} \lsim m_N \lsim 1\GeV$.

Having established the viability of this model in the broad cosmological context, there are several further issues which could be addressed. A detailed investigation of various astrophysical probes for the DM candidate should be carried out.  Also, as we have emphasized, the scalar sector of the model can also lead to strong first order electroweak phase transition, which is a basic requirement for successful baryogenesis. We leave the more detailed investigation of these issues within this model for future work.

%
\section*{Acknowledgements}
\label{sec:acknowledgements}
%

We acknowledge the financial support from the
Academy of Finland, projects 278722 and 267842. VV acknowledges the financial support from the Finnish Cultural foundation and Magnus Ehrnrooth foundation.

%
\appendix
%

%
\section{Cross sections}
\label{xsect}
%

Here we give the formulas for the computation of the annihilation cross section for the model considered in Sec. \ref{model}. To make the equations more concise, it is useful to define the couplings, which at the electroweak broken vacuum read
\begin{align}
\begin{split}
\lambda_\mathit{HHH} &= -6 v \lambda_\phi \cos^3\beta - 3 \muphis \cos^2\beta \sin\beta - 3 v \lphis \cos\beta \sin^2\beta - 2 \mu_3 \sin^3\beta, 
\end{split} \\
\begin{split}
\lambda_\mathit{HHS} &= 6 v \lambda_\phi  \cos^2\beta \sin\beta - \muphis \cos^3\beta + 2 \muphis \cos\beta \sin^2\beta \\
&- 2 v \lphis \cos^2\beta \sin\beta + v \lphis \sin^3\beta - 2 \mu_3 \sin^2\beta \cos\beta, 
\end{split}\\
\begin{split} \label{hHHcoup}
\lambda_\mathit{HSS} &= -6 v \lambda_\phi  \cos\beta \sin^2\beta + 2 \muphis \cos^2\beta \sin\beta - v \lphis \cos^3\beta \\
&- \muphis \sin^3\beta+2 v \lphis \sin^2\beta \cos\beta - 2 \mu_3 \sin\beta \cos^2\beta, 
\end{split}\\
\begin{split}
\lambda_\mathit{SSS} &= 6 v \lambda_\phi  \sin^3\beta - 3 \muphis \sin^2\beta \cos\beta + 3 v \lphis \sin\beta \cos^2\beta - 2 \mu_3 \cos^3\beta, 
\end{split}
\end{align}
and
\bea
g_{s,H} = g_s \sin\beta, \quad g_{p,H} = g_p \sin\beta, \quad g_{s,S} = g_s \cos\beta, \quad g_{p,S} = g_p \cos\beta.
\eea
Decay widths for the processes $H\to\psi\overline\psi$ and $S\to\psi\overline\psi$ are
\be
\Gamma_{X\to\psi\overline\psi}=\Theta(m_X-2m_\psi) \frac{m_X R_X^2}{8\pi} \left(g_s^2 \left(1-\frac{4m_\psi^2}{m_X^2}\right)^\frac{3}{2} + g_p^2 \sqrt{1-\frac{4m_\psi^2}{m_X^2}} \right).
\ee
where $X = H,S$ and $R_H = \sin\beta$, $R_S = \cos\beta$. Cross section for the process $ab \to cd$ reads
\be
\sigma(ab \to cd) = \frac{1}{16 \pi \lambda(s,m_a^2,m_b^2)} \int_{t_-}^{t_+} \mathrm{d}t \left|T(ab \to cd)\right|^2,
\ee
where $\lambda(x,y,z) = x^2 + y^2 + z^2 - 2xy - 2xz - 2yz$ and 
\bea
t_\pm =& m_a^2 + m_c^2 \pm \frac{\sqrt{\lambda(s,m_a^2,m_b^2)\lambda(s,m_c^2,m_d^2)}}{2s} \\
&\pm 2\sqrt{\left(m_a^2 + \frac{\lambda(s,m_a^2,m_b^2)}{4s}\right)\left(m_c^2 + \frac{\lambda(s,m_c^2,m_d^2)}{4s}\right)}.
\eea
The squared amplitude, averaged over the initial states and summed over the final states, for the DM, $\psi$, annihilating to two scalars, $h_i,h_j = H,S$, or equivalently for $h_i$ and $h_j$ annihilating to $\psi\overline\psi$ -pair is given by
\bea
\left|T_{ij}\right|^2 &= \left(1-\frac{\delta_{ij}}{2}\right)\left(T_{ij}^{(ss)} + T_{ij}^{(tt)} + T_{ij}^{(uu)} + 2 T_{ij}^{(st)} + 2 T_{ij}^{(su)} + 2 T_{ij}^{(tu)}\right),
\eea
where
\begin{align}
\begin{split}
T_{ij}^{(ss)} =& \left(2-\delta_{ij}\right) \left((s-4m_\psi^2)\left|\sum_{k=H,S} \frac{\lambda_{ijk} g_{s,k}}{s-m_k^2 + \mathrm{i} m_k \Gamma_k}\right|^2  \right.\\&\left. + s \left|\sum_{k=H,S} \frac{\lambda_{ijk} g_{p,k}}{s-m_k^2 + \mathrm{i} m_k \Gamma_k}\right|^2 \right) ,
\end{split} \\
\begin{split}
T_{ij}^{(tt)} =&\left( 2 \left(m_\psi^4-m_\psi^2 (t+u)-m_i^2 m_j^2+t u\right) g_{s,i}^2 g_{s,j}^2 \right.\\ 
&\left. +2 \left(m_\psi^4+m_\psi^2 (u-3 t)-m_i^2 m_j^2+t u\right) \left(g_{s,i}^2 g_{p,j}^2 + g_{p,i}^2 g_{s,j}^2\right) \right.\\ 
&\left. -2 \left(7 m_\psi^4+m_\psi^2 (-4 m_i^2-4 m_j^2+9 t+u)+m_i^2 m_j^2-t u\right) g_{p,i}^2 g_{p,j}^2 \right.\\ 
&\left. +8 m^2 s g_{s,i}g_{s,j}g_{p,i}g_{p,j} \right) (t-m_\psi^2)^{-2} ,
\end{split}\\
\begin{split}
T_{ij}^{(uu)} =&\left( 2 \left(m_\psi^4-m_\psi^2 (t+u)-m_i^2 m_j^2+t u\right) g_{s,i}^2 g_{s,j}^2 \right.\\ 
&\left. +2 \left(m_\psi^4+m_\psi^2 (t-3 u)-m_i^2 m_j^2+t u\right) \left(g_{s,i}^2 g_{p,j}^2 + g_{p,i}^2 g_{s,j}^2\right) \right.\\ 
&\left. -2 \left(7 m_\psi^4+m_\psi^2 (-4 m_i^2-4 m_j^2+9 u+t)+m_i^2 m_j^2-t u\right) g_{p,i}^2 g_{p,j}^2 \right.\\ 
&\left. +8 m^2 s g_{s,i}g_{s,j}g_{p,i}g_{p,j} \right) (u-m_\psi^2)^{-2} ,
\end{split}\\
\begin{split}
T_{ij}^{(st)} =& \sum_{k=H,S}  \frac{\lambda_{ijk}}{\left|s - m_k^2 + \mathrm{i} m_k \Gamma_k \right| (t - m_\psi^2)} \\
&\times\left( \left( 2 m_\psi (t-u) g_{s,i} g_{s,j} + 2 m_\psi (4 m_\psi^2-2 m_i^2-2 m_j^2+3 t+u) g_{p,i} g_{p,j} \right) g_{s,k} \right.\\
&\left.+ \left( 2 m_\psi (m_i^2-m_j^2+s) g_{s,i} g_{p,j} + 2 m_\psi (-m_i^2+m_j^2+s) g_{p,i} g_{s,j} \right) g_{p,k} \right) ,
\end{split}\\
\begin{split}
T_{ij}^{(st)} =& \sum_{k=H,S}  \frac{\lambda_{ijk}}{\left|s - m_k^2 + \mathrm{i} m_k \Gamma_k \right| (u - m_\psi^2)} \\
&\times\left( \left( 2 m_\psi (u-t) g_{s,i} g_{s,j} + 2 m_\psi (4 m_\psi^2-2 m_i^2-2 m_j^2+3 u+t) g_{p,i} g_{p,j} \right) g_{s,k} \right.\\
&\left.+ \left( 2 m_\psi (m_i^2-m_j^2+s) g_{s,i} g_{p,j} + 2 m_\psi (-m_i^2+m_j^2+s) g_{p,i} g_{s,j} \right) g_{p,k} \right) ,
\end{split}\\
\begin{split}
T_{ij}^{(tu)} =&\left( 2 \left(m_\psi^4-m_\psi^2 (t+u)-m_i^2 m_j^2+t u\right) g_{s,i}^2 g_{s,j}^2 \right.\\ 
&\left. -2 \left(m_\psi^4+m_\psi^2 (t-3 u)-m_i^2 m_j^2+t u\right) g_{s,i}^2 g_{p,j}^2 \right.\\ 
&\left. -2 \left(m_\psi^4+m_\psi^2 (u-3 t)-m_i^2 m_j^2+t u\right) g_{p,i}^2 g_{s,j}^2 \right.\\ 
&\left. -2 \left(9 m_\psi^4+m_\psi^2 (-4 m_i^2-4 m_j^2+3(t+u))-m_i^2 m_j^2+t u\right) g_{p,i}^2 g_{p,j}^2 \right.\\ 
&\left. +8 m^2 s g_{s,i}g_{s,j}g_{p,i}g_{p,j} \right) (t-m_\psi^2)^{-1} (u-m_\psi^2)^{-1} .
\end{split}
\end{align}

%
\bibliography{pseudo.bib}
%

\end{document}